\documentstyle[12pt]{article}
\begin{document}
\newcommand{\newc}{\newcommand}
\newc{\zeus}{{\sc Zeus}}
\newc{\gev}{\,GeV}
\newc{\rp}{$R_p$}
\newc{\rpv}{$\not\!\!R_p$}
\newc{\rpvm}{{\not\!\! R_p}}
\newc{\rpvs}{{\not R_p}}
\newc{\ra}{\rightarrow}
\newc{\lsim}{\buildrel{<}\over{\sim}}
\newc{\gsim}{\buildrel{>}\over{\sim}}
\newc{\esim}{\buildrel{\sim}\over{-}}
\newc{\lam}{\lambda}
\newc{\lsp}{{\tilde\Lambda}}
\newc{\oc}{{\cal {O}}}
\newc{\msq}{m_{\tilde q}}
\newc{\mpl}{M_{Pl}}
\newc{\mw}{M_W}
\newc{\pho}{{\tilde\gamma}}
\newc{\half}{\frac{1}{2}}
\newc{\third}{\frac{1}{3}}
\newc{\quarter}{\frac{1}{4}}
\newc{\beq}{\begin{equation}}
\newc{\eeq}{\end{equation}}
\newc{\barr}{\begin{eqnarray}}
\newc{\earr}{\end{eqnarray}}
\newc{\ptmis}{\not\!\!{p_T}}
\newc{\mc}{\multicolumn}
\addtocounter{table}{1}

\title{ R-Parity Violation at HERA}
\author{J. Butterworth and H.~Dreiner}
\date{{\small Department of Physics, University of Oxford,\\
  1 Keble Rd, Oxford OX1 3RH}}
\maketitle

\begin{abstract}

\noindent We summarize the signals at HERA in supersymmetric  models with
explicitly broken R-parity. As the most promising case, we consider  in detail
the resonant production of single squarks through an operator $L_1Q_i{ \bar
D}_j$, a production process analogous to that for leptoquarks. However, the
dominant decay of the squark to a quark and a photino leads to a very different
experimental signature. We examine in particular the  case where the photino
decays to a positron and two quarks. Using a detailed Monte-Carlo procedure we
obtain a discovery limit in the squark  mass---Yukawa coupling plane. HERA can
discover a squark for a mass as large as $270 \gev$ and for an R-parity
violating Yukawa coupling as small as $5.8 \times 10^{-3}$.

\end{abstract}

\section {Introduction}
Supersymmetry \cite{susy} (SUSY) is a promising solution to the gauge-hierarchy
problem \cite{hierarchy}. It has been thoroughly analysed in the minimal
supersymmetric standard model (MSSM), which has the minimal field content {\it
and} the minimal superpotential consistent with the Standard Model (SM). The
MSSM has a discrete, multiplicative symmetry known as R-parity
\beq
R_p=(-)^{2S+3B+L},
\label{eq:rp}
\end{equation}
where $S,B,$ and $L$ are the spin, the baryon and the lepton quantum numbers,
respectively. The fields of the SM, including the extra scalar Higgs doublet,
have $R_p=+1$; their superpartners have $R_p=-1$. Recently there has been much
theoretical \cite{ibanross}, cosmological
\cite{rpcosmo1,rpcosmo2,bruceII,baryog} and
phenomenological work \cite{butter,rptop,dp,rplola} on the more general
supersymmetric extension of the SM with explicitly broken R-parity (\rpv). As
in the MSSM, this only requires the minimal field content consistent with the
SM, but in contrast allows for {\it all} gauge and supersymmetric invariant
terms in the superpotential including now \rpv\ terms which violate either
lepton- or baryon-number. \rpv\ has been shown to be theoretically consistent
with grand unification theories \cite{rp1,flipped}, as well as with superstring
theory \cite{rpstring}, and it is possibly better theoretically motivated than
the MSSM \cite{ibanross}. There has also been an extensive discussion on
cosmological bounds on {\it all} \rpv-operators \cite{rpcosmo1,bruceII,baryog}.
In Ref.\cite{baryog} it was shown that most bounds no longer apply, given
reasonable assumptions on the mechanism for mattergenesis. We therefore find it
very much worthwhile to experimentally search for \rpv.

To date the most stringent experimental bounds on the \rpv\ Yukawa couplings
have been determined from virtual effects of the new supersymmetric states in
processes involving only SM final states. \rpv\ could be found, or these
bounds substantially improved, by direct searches for supersymmetric particles
at colliders.

The single squark production mechanism is analogous to that of leptoquarks.
Thus it is important to discriminate \rpv\ squark from leptoquark production.
We study in detail the distinguishing dominant cascade decay of the squarks to
a quark and a photino and show that searching for this decay leads to the
greatest discovery potential in the squark mass--Yukawa coupling plane.

In the following section we will give a brief theoretical overview of \rpv\
models. Then, in section~3 we will procede to discuss the prospects for
discovering supersymmetry at HERA, both for the MSSM and for \rpv\ models.  We
show that, while HERA is conventionally considered to be relatively
uncompetetive with present limits for supersymmetric searches in the
MSSM~\cite{chll,ruckl,heraws} , it is well suited to study a subclass of
\rpv-models, which allow for resonant single squark production. Identifying the
most promising case, in section~4 we then present cross-section and branching
fraction calculations. Using these as input to a detailed Monte-Carlo of these
processes, and coupling this with a simulation of the \zeus\ detector, we
examine possible backgrounds to these processes and evaluate the discovery
potential in \zeus. We show that \rpv\ leads to a much larger mass reach than
in MSSM searches and gives HERA the possibility of finding supersymmetry.

\section{Theoretical Overview}
Given the particle content of the SM, the most general gauge and supersymmetry
invariant superpotential is given by \cite{gaugterms}
\beq
W = W_{MSSM} + W_{\rpvs},
\label{eq:superpot}
\end{equation}
where the first term includes those couplings which give mass to the matter
fields and which are included in the MSSM. The second term in
Eq.(\ref{eq:superpot}) consists of the \rpv\  Yukawa couplings\footnote{For a
review of supersymmetry and in particular the notation employed here see
Ref.\cite{susy}.}
\beq
W_{\rpvs } =  \lam_{ijk} [L_{i}L_{j}\bar{E}_{k}]_F + \lam'_{ijk} [L_{i}
Q_{j}  \bar{D}_{k}]_F  + \lam''_{ijk} [\bar{U}_{i} \bar{D}_{j}\bar{D}_{k}]_F.
\label{eq:operators}
\eeq
Here we have used the superfield notation; $L$ and $\bar{E}$ ($Q$ and
\(\bar{U}, \bar{D}\)) are the (left-handed) lepton doublet and the antilepton
singlet (quark doublet and antiquark singlet) chiral superfields respectively.
$\lam, \lam'$ and $\lam''$ are dimensionless coupling constants and $i,j,k$ are
generation indices. $SU(2)$ and $SU(3)$ indices have been suppressed here and
throughout. The subscript $F$ denotes the supersymmetric invariant part in the
product of chiral superfields.

As well as breaking R-Parity, the first two sets of terms in
Eq.(\ref{eq:operators}) violate lepton number conservation, and the last set
violates baryon number conservation. Together they can lead to an unacceptable
level of proton decay, a problem which is avoided if $R_p$ conservation is
imposed ($W_{\rpvs}\equiv0$). However, it is sufficient to impose a discrete
symmetry which allows for either only the lepton-number violating terms
($LL{\bar E}$, $LQ{\bar D}$) or for only the baryon number violating terms
(${\bar U}{\bar D}{\bar D}$) in the Lagrangian.\footnote{As discussed in
Ref.\cite{baryog} it is sufficient to exclude the simultaneous presence of the
$[LQ {\bar D}]_F$ terms and the $[{\bar U}{\bar D}{\bar D}]_F$ terms.}  These
discrete symmetries are theoretically on an equal footing with each other and
with $R_p$ \cite{baryog} with the one allowing for the lepton-number violating
\rpv-operators possibly preferred \cite{ibanross}.

At HERA we are particularly interested in the case where an operator $[L_1Q_i
{\bar D}_j]_F$ is dominant, since this leads to resonant squark production. We
therefore present the Yukawa couplings explicitly in terms of the 4-component
Dirac states
\barr
{\cal L}_{L_1Q_i{\bar D}_j} &=& \lam'_{1ij} \left[ -{\tilde e}_L u_L^i
{\bar d}_R^j -e_L{\tilde u}_{L}^i {\bar d}_R^j - ({\bar e}_L)^c u_L^i {\tilde
d}_R^j \right. \nonumber \\
&& \left. +{\tilde \nu}_L d_L^i {\bar d}_R^j
+ \nu_L{\tilde d}_{L}^i {\bar d}_R^j + ({\bar \nu}_L)^c d_L^i {\tilde d}_R^j
\right] + h.c.
\label{eq:lqdlagrangian}
\earr
Here the tilde denotes the scalar superpartners and the superscript $^c$
denotes
the charge conjugated spinor. The minus signs are due to the $SU(2)$ structure
of the term.

\section{Supersymmetry at HERA}

In discussing the possible \rpv\ signals at HERA we shall closely follow the
discussion and notation of Ref.\cite{rphadron}. In particular we shall also
make the following simplifying assumptions \begin{itemize}

\item
We consider only those cases where in turn one of the 45 operators in
Eq.(\ref{eq:operators}) dominates and the others are negligible, {\it i.e.}
there is a hierarchy in the \rpv\ Yukawa couplings. This is a reasonable
assumption since the known Yukawa couplings to the Higgs in the SM show such a
hierarchy ($m_ {elec}/m_{top}\simeq 3 \cdot 10^{-6}$).

\item

The lightest supersymmetric particle (LSP) is a neutralino, which we denote by
$\lsp$. This is a non-trivial assumption, since the cosmological bounds, which
restrict the LSP to be a neutralino in the MSSM \cite{susydm} no longer apply
in the case of \rpv\ \footnote{See Ref.\cite{bruceII} for a brief discussion of
relic LSP densities in the case of \rpv}. However, renormalization group
analyses in minimal supergravity models, which shouldn't be much affected by a
small \rpv\  Yukawa coupling, suggest that the LSP is indeed a neutralino
\cite{ibanlop}. When studying specific signals in detail we will for
simplification assume the LSP to be the photino.

\end{itemize}

The phenomenology of \rpv\ differs from that of the MSSM in two main aspects:
(1) the LSP is no longer stable since it is not protected by a symmetry; it can
therefore decay within the detector; and (2) it is possible to have single
production of supersymmetric particles, since the final state is no longer
restricted to be $R_p$-even. We shall make use of both of these points when
discussing the dominant \rpv-signal at HERA in section \ref{sec:signal}.

\medskip

(1) The LSP can decay {\it via} the process shown in Figure~1, where we have
circled
the \rpv\  vertex. It will decay either within or outside of the detector. The
condition that the LSP decays within the detector translates into a lower bound
on the \rpv-Yukawa coupling \cite{dawson,rphadron}
\beq
\lam_{\,\rpvm} > 1.6\gamma\cdot10^{-4} (m_{{\tilde f}}/ 100~\gev)^2 (10\gev/
M_{\lsp} )^{5/2},
\label{eq:lam}
\eeq
where $\gamma$ is the Lorentz boost factor in the lab-frame. $m_{{\tilde f}}$
is the mass of the scalar fermion which enters via the propagator in the LSP
decay ({\it cf.} Figure 1). Here we have made the approximations of massless
final state particles and $m_{{\tilde f}}\gg M_{\lsp}$, and demanded that
$c\gamma\tau_{LSP} \lsim 1m$, where $\tau_{LSP}$ is the mean lifetime of the
LSP. The constraint (\ref{eq:lam}) is always satisfied for the ranges of the
squark mass and the \rpv-Yukawa coupling which HERA can probe through single
squark production, provided $M_\lsp \gsim 10\gev$.

(2) In \rpv\ models there are two dominant production mechanisms for
supersymmetric particles. First, there is the conventional pair or associated
production through MSSM (super-) gauge couplings. In addition, there is also
the possibility of single scalar fermion production directly through
\rpv-couplings, which is {\it not} possible in the MSSM. In both cases the
supersymmetric states can decay within the detector to lower mass, $R_p$-even
states giving the two sets of signals
\barr
({\rm SUSY \,\,Pair }\ {\rm production}) & {\it with} &(\rpvm  -{\rm decay}),
\label{eq:signal1} \\
(\rpvm,  \,{\rm single}\ {\rm SUSY}\ {\rm production}) & {\it with} & (\rpvm
-{\rm decay}),
\label{eq:signal2}
\earr
which we have summarized in Table~1. The table is analogous to the table
presented in Ref.\cite{rphadron}; for completeness we discuss it in detail in
the appendix. Throughout we assume that the inequality (\ref{eq:lam}) is
satisfied and thus assume all final states to be \rp\ even. If the LSP does
{\it
not} decay within the detector, the signal (\ref{eq:signal1}) reduces to the
standard scenario in the MSSM with missing $p_T$ ($\ptmis$)  signals which have
been widely discussed. For a long-lived LSP the process (\ref{eq:signal2})
could
still lead to novel signals. However, the cross section for single SUSY
production is proportional to $(\lam' )^2$ and thus strongly suppressed in the
case where the Yukawa coupling $\lam'$ violates the inequality (\ref{eq:lam}).
In section \ref{sec:mssmprod} we discuss the case (\ref{eq:signal1}) and in the
remainder of the paper we discuss the  more promising case (\ref{eq:signal2})
in
detail. However, before moving on we review the possibilities for observing
supersymmetry in the case of the MSSM (section~\ref{sec:mssmdec}).

\subsection{MSSM at HERA}
\label{sec:mssmdec}
In the case of conserved $R_p$, the LSP is stable. At HERA, the two dominant
production mechanisms of the MSSM are \cite{chll,ruckl,heraws}
\barr
e^-+q &\ra& {\tilde \nu}_e + {\tilde q}, \label{eq:mssmch} \\
e^-+q &\ra& {\tilde e}^- + {\tilde q}.  \label{eq:mssmn}
\earr
In the reaction (\ref{eq:mssmn}), the decay products are neutral in most
scenarios and therefore hard to detect. We focus on the neutral current process
(\ref{eq:mssmn}); for an integrated luminosity of $200\,pb^{-1}$, the
discovery potential is \cite{ruckl}\footnote{A more optimistic discovery reach
was optained in Ref.\cite{montag}, however not within the MSSM since the
neutralino masses were varied independently of each other.}
\beq
m_{{\tilde e}}+ \msq \lsim 200\gev.
\label{eq:mssmreach}
\eeq
In obtaining this limit the authors assumed that there are no decays of
the squark or selectron other than that to the LSP.

The discovery reach is to be compared with the existing experimental scalar
fermion mass bounds. From LEP we have the model independent bounds
\cite{lepbound}
\beq
m_{{\tilde e}},\, \msq \gsim 45\gev, \label{eq:lep}
\eeq
which are determined from the $Z^0$ width. CDF quote a much stronger, but model
dependent bound for the squark mass \cite{cdfbound}
\beq
\msq \gsim 150 \gev,
\label{eq:cdfq1}
\eeq
for which it was also assumed that there are no cascade decays of the squarks
and that all six squarks are degenerate in mass. In contrast, if one assumes a
spectrum of non-degenerate squark masses, with one squark substantially lighter
than the others, the bound on the lightest squark is much  weaker
\cite{manuel-stop}, $\msq \gsim 64\gev$.  Comparing the experimental bounds
with the discovery reach at HERA (\ref{eq:mssmreach}) we can conclude that for
6 degenerate squarks there is only a very small window of discovery for the
MSSM. If the top squark is much lighter than the other squarks
\cite{lightstop}, then there is still the possibility of discovering $R_p$
conserving supersymmetry at HERA \cite{herastop}.

\subsection{Pair Production with \rpv\  Decay}
\label{sec:mssmprod}

In \rpv models, the LSP decays and the standard $\ptmis$ signal is either
diluted or completely absent, depending on the dominant \rpv-operator. The
decays of the supersymmetric particles in the various \rpv-models have been
discussed in detail in Ref.\cite{rphadron}. We have made use of these results
in compiling the dominant signals of the processes (\ref{eq:signal1}) in the
first three rows of Table~1.

For a dominant operator $[L_iL_j{\bar E}_k]_F$, for example, the LSP decays to
two charged leptons and a neutrino,
\beq
\lsp  \ra  e^\pm_i + e^\mp_k + \nu_j, \quad [L_iL_j{\bar E}_k]_F ,
\label{eq:lle}
\eeq
which in the case of the charged current production (\ref{eq:mssmch}) results
in
a signal of 5 charged leptons! (the entry in the first row and the second
column). In the case of a dominant operator $[L_iQ_j{\bar D}_k]_F$ we have
\beq
\lsp\ra (e^\pm_i,\nu_i,{\bar\nu}_i) + 2\,{\rm jets}, \quad [L_iQ_j{\bar
D}_k]_F,
\label{eq:lqd}
\eeq
and the optimal signal still contains 3 charged leptons.

{}From Table~1 we conclude, that in the case of pair production and
\rpv-decays,
the dominant operator $[{\bar U}{\bar D}{\bar D}]_F$ does not lead to any
promising signals. The two most promising signals for discovery of SUSY are the
charged current production (\ref{eq:mssmch}) followed by the LSP decays
(\ref{eq:lle}) or (\ref{eq:lqd}).

In order to get a rough estimate of the discovery potential at HERA for these
two models we make use of the total production cross section $\sigma(e^-P\ra
{\tilde e}^-{\tilde q}+X)$ taken from reference~\cite{ruckl}. For the dominant
operator $[L_iL_j{\bar E}_k]_F$ we restrict ourselves to first and second
generation charged leptons: $i,j,k=1,2$; there is no SM background for the
production of five charged leptons in the final state. As an optimistic
estimate, we demand a discovery limit of five events, before any acceptance
cuts, for an integrated luminosity of $200\, pb^{-1}$ to be sufficient for
detection. Taking the most optimistic scenario for the neutralino sector of
Figure~2, Ref \cite{ruckl}, we thus obtain a maximal reach of
\beq
m_{{\tilde e}} + \msq \leq 216 \gev,
\quad [L_iL_j{\bar E}_k],\,\,i,j,k=1,2.
\label{eq:reachlle}
\eeq
Next we consider the case (\ref{eq:lqd}). For three charged leptons
($e$ or $\mu$),
the production rate is suppressed by a branching fraction squared $(0.88)^2$
({\it cf.} Section \ref{sec:decrates}). If we furthermore require like-sign
tri-leptons the production rate is instead reduced by a factor of $(0.44)^2$.
Taking this latter case, which eliminates all SM background, and
again optimistically only requiring five observed events for an integrated
luminosity of $200\, pb^{-1}$ we obtain as a rough estimate of the discovery
limit\footnote{We note that for 3 charged leptons in the final state there is
the physical background from $Z^0$ production, $e^-+p\ra e^-+Z^0+X$, followed
by
the decay $Z^0\ra (ee, \mu\mu)$. The total cross section multiplied by the
relevant branching fraction to two charged leptons $(e,\mu)$ is $\sigma\simeq
2.0\cdot10^{-2}\,pb$, Ref.\cite{zep2}. About $50\%$ of the scattered electrons
vanish down the beam pipe and thus the background is roughly a factor five
smaller than the signal we are interested in (with no explicit lepton-number
violation), and should be included in a proper background study.}
\beq
m_{{\tilde e}}+ \msq \leq 165\gev,\quad L_iQ_j{\bar D}_k,\,\, i\not=3.
\label{eq:reachlqd}
\eeq
We now compare these discovery limits with the experimental mass bounds. The
model independent LEP bounds (\ref{eq:lep}) still hold in the case of \rpv. The
CDF bounds on the squark masses are not valid  since they relied on $\ptmis$
signals. However, using the CDF di-lepton search data an independent stringent
bound on the squark masses has been determined, which incorporates \rpv\ decays
of the supersymmetric particles \cite{dp}
\barr
\msq &\geq&175\gev,\qquad [L_iL_j{\bar E}_k],\,\,\,
i,j,k=1,2, \label{eq:sqlle2}  \\
\msq & \geq & 100\gev,\qquad [L_iQ_j{\bar D}_k],\,\,i,j\not=3.
\label{eq:sqlqd}
\earr
In deriving these bounds it was assumed that the LSP is the photino and that
there are {\it five} mass degenerate squarks, thereby allowing for the
possibility of a light top squark. Furthermore it was assumed that the
branching
fraction $BF({\tilde q}\ra q+\pho)=100\%$. Using the CDF di-lepton data the
mass
bound for a single light squark is as low as the LEP bound \cite{rptop}, again
assuming $BF({\tilde q}\ra q+\pho)=100\%$. For $BF({\tilde q}\ra
e^\pm+q')=50\%$
and $BF({\tilde q}\ra q+\pho)=0$, which is unlikely, $\msq \geq 108\gev$
\cite{rptop}, so that a realistic bound for a light squark is presumably
somewhat stronger than $45\gev$.

\medskip

{}From our rough estimates we conclude that for five degenerate squarks there
is
no discovery window at HERA for a dominant operator $L_iL_j{\bar E}_k$. In the
case of an operator $L_iQ_j{\bar D}_k$ there is still a small discovery window
of about $20\gev$, which warrants a proper analysis for $i=2,3$ and/or $j=3$.
For $i=1$, the \rpv\ single squark production at HERA leads to a significantly
larger discovery potential than the pair production. In sections (4) and (5) we
discuss this in detail.

\subsection{\rpv-Production with \rpv-Decay}
\label{sec:signal}

At HERA, \rpv\ includes the unique possibility of single squark production
through the couplings given in Eq.(\ref{eq:lqdlagrangian}), and the diagram
shown in Figure~2a. We have summarized the possible signals for single squark
production in the last 2 rows of Table~1. Single squark production has
previously been discussed in Refs.\cite{herastop,joanne}. In both references
only the \rpv-decays of the squarks, as in Figure~2b are studied and in
Ref.\cite{herastop} the discussion is limited to the production of top squarks.
We extend these analyses to include the dominant cascade decays of the squarks,
${{ \tilde q}}\ra q+\pho$, followed by the \rpv-decay of the photino, as in
Eq.(\ref{eq:lqd}) and shown in Figure~2c. Throughout this discussion we make
the simplifying assumption that the LSP is the photino, {\it i.e} it has no
Zino or Higgsino admixture.

\subsubsection{Analogies to Leptoquark Production}

The single production of squarks at HERA is analogous to the production of
leptoquarks. In the notation of Ref.\cite{brw} the ${\tilde u}_i$ squark
corresponds to the ${\tilde R}_2$ leptoquark, and the ${\tilde {\bar d} }_j$
squark corresponds to the $S_1$ leptoquark.\footnote{The quantum number $F$ of
Table~1 in Ref.\cite{brw} does not apply to the corresponding squarks since it
was computed assuming the baryon- and lepton-number are conserved at the
production vertex.} The squarks can decay back to the initial state ($e^-q$)
through the Yukawa coupling (\ref{eq:lqdlagrangian}) just as the leptoquarks
do, and as shown in Figure~2b. This is observable as a resonant peak in the
neutral current deep inelastic differential cross section $d^2\sigma(e^-p\ra
e^-X)/dxd Q^2, $ \footnote{For the ${\tilde{\bar d}}_j$ case this is also
observable in the charged current cross section.} and we can translate most
results on the $S_1$ and ${\tilde R}_2$ leptoquarks into results on \rpv. In
particular from Ref.\cite{brw}, Eq.(8) one expects to be able to observe a
down-like squark for $m_{{\tilde d}}\simeq200 \gev$ and a Yukawa coupling as
small as $\lam'\simeq 0.01$. This reach is significantly larger than that of
Eq.(\ref{eq:reachlqd}), where the pair production was followed by $LQ{\bar D}$
mediated decays.

\medskip

It is important to discriminate between the discovery of a leptoquark and of
supersymmetry with \rpv. There are several suggestions as to how this might be
done.

\medskip

(1)~As can be seen from the couplings in Eq.(\ref{eq:lqdlagrangian}), it is
only the left-handed electrons which couple to the squarks, whereas some
leptoquarks can couple to both left- and right-handed electrons. Thus using
polarized beams it could be possible to distinguish these two possibilities
\cite{joanne}. However, if the right-handed coupling of the relevant leptoquark
vanishes, as it very well may, then this method does not lead to a
distinguishing signal.

\medskip

(2) In Ref.\cite{herastop} it was suggested that one could distinguish up-like
squarks from leptoquarks through their different decay modes which are mediated
by the Yukawa couplings. Through the couplings (\ref{eq:lqdlagrangian}) the
${\tilde {u_i}}$ can only decay as ${\tilde{{\bar u}_i}}\ra e^-+{\bar d}_{jR}$,
whereas the $S_1$ leptoquark has two decay modes $S_1\ra e^-+u,\,\, \nu + d$
and thus also contributes to the {\it charged} current deep inelastic
scattering. However as we pointed out above, there are leptoquarks other than
$S_1$. For  example, ${\tilde R}_2$ has identical Yukawa-couplings to the
up-like squark  and also decays as ${\tilde R}_2\ra e^-+{\bar d}_{jR}$.
Therefore this procedure can not be used to distinguish a squark from a
leptoquark.

\medskip

(3) In supersymmetric theories the squarks have specific gauge couplings
through
which they can cascade decay to the LSP, ${{\tilde q}}\ra q+\pho$. We study
these decays as a characteristic feature of squarks and thus study the
processes\footnote{To our knowledge this was first proposed in
Ref.\cite{joanne}.}
\barr
e^-+ u_i & \ra ({\tilde d}_{jR})^* &\ra d_{jR}+\pho,
\label{eq:one} \\
e^-+ {\bar d}_j  & \ra ({\tilde{\bar u}}_{iL})^* &\ra {\bar u}_i +\pho ,
\label{eq:two}
\earr
which are shown in Figure~2c. The asterisk indicates that the produced squark
can be virtual. The photino decays as in Eq.(\ref{eq:lqd}), and we focus on the
decay $\pho\ra e^++2$ jets. This leads to a single {\it positron} in the final
state, three jets, and no electron or significant $\ptmis$. This strongly
reduces the background compared to the other photino decays but also introduces
a branching fraction of about $0.44$ ({\it cf.} Section \ref{sec:decrates}).
The case of the decay $\pho\ra\nu_e+2$ jets leads to a weak $\ptmis$ signal
with large background, which we do not consider promising and have thus also
not included in Table~1.

In the following we assume that the only relevant decays of the produced
squarks
are given by
\barr
{{{\tilde d}_{jR}}} &\ra& \left\{ \begin{array}{l} e^-_L+u_{iL} \\ \nu_e +
d_{jR} \\ d_{jR} + \pho \end{array} \right. , \label{eq:ddec} \\
{\tilde{\bar u}}_{iL} &\ra& \left\{ \begin{array}{l} e^-_L + {\bar d}_{jR} \\
{\bar u}_{iL} + \pho \end{array} \right. , \label{eq:udec}
\earr
which we discuss in more detail in section \ref{sec:decrates1}. Thus we assume
that there are no further cascade decays for example to other neutralinos or to
the gluino.

\medskip

(4) A direct consequence of Eqs.(\ref{eq:ddec},\ref{eq:udec}) is that the total
decay width of the squarks should  differ from the corresponding decay widths
of
the leptoquarks. In neutral current deep-inelastic scattering one can in
principle measure the mass, coupling and decay width of a ``leptoquark"
resonance. For a true leptoquark the decay width is given in terms of
the mass and one coupling. If the observed resonance is a squark one should
find
the measured width to be larger than the calculated leptoquark width by up to
an
order of magnitude. However, it is unlikely that the experimental resolution
will be sufficient to accomplish this \cite{tariq}.

\subsubsection{Existing Limits}

Before moving on to discuss the processes (\ref{eq:one},\ref{eq:two})
in detail we first consider existing experimental bounds on the Yukawa
couplings $\lam'_{1ij}$, which lead to resonant squark production at HERA. In
our study we limit ourselves to the couplings
\beq
\lam'_{1ij}, \quad i=1,2;\,\,\,j=1,2,3.
\eeq
We do not consider the case where $i=3$, as this implies a top quark in the
initial and/or final state. The case of $i=3$ has been discussed in
Ref.\cite{herastop}, where the authors suggest that the squarks could be
discovered {\it via} the leptoquark-analogous \rpv-decay. However, in that case
it is not clear how the squark could be distinguished from a leptoquark.

\medskip

The squarks and sleptons can contribute virtually to various processes with
\rp-even final states through the relevant \rpv-operators. The experimental
values or bounds of these processes directly lead to limits on any contribution
beyond that of the SM. In our case this leads to bounds on the \rpv-Yukawa
couplings as a function of the squark masses.

\medskip

The operators $L_1Q_1{\bar D}_j$, $j=1,2,3$, contribute to the semileptonic
decays of quarks \cite{vernon}. The $(V-A)\otimes(V-A)$ structure of the \rpv,
effective four-fermion charged current interaction is identical to that of the
SM. Therefore its contribution is equivalent to a shift in the Fermi-constant.
If, as we have assumed, {\it one} \rpv-operator is dominant, then this is a
violation of charged current universality \cite{vernon}. The experimental
limits lead to a bound on the corresponding Yukawa coupling at the {\it two}
sigma level
\beq
\lam'_{11j} \lsim 0.03 \left( \frac{m_{{\tilde d}_{jR}}}{100\gev} \right) ,
\quad j=1,2,3,
\label{eq:bound1}
\eeq
and we note that the bound is only correlated to the $SU(2)$ singlet, down-like
squark mass, not the doublet, up-like squark mass. If we allow for more than
one
\rpv\ operator to contribute, then the violation of universality is reduced and
the above bound is correspondingly weakened.

The operators $L_1Q_2{\bar D}_j$, $j=1,2,3,$ contribute to forward-backward
asymmetries measured in $e^+e^-$-collisions. Requiring the combination of \rpv\
and SM contribution to agree with the experimental data at the one sigma level
one obtains bounds on the Yukawa coupling constants \cite{vernon}
\barr
\lam'_{12j} &\lsim & 0.45 \left( \frac{m_{{\tilde d}_{jR}}}{100\gev}\right),
\quad j=1,2,3, \\
\lam'_{123} &\lsim & 0.26  \left( \frac{m_{{\tilde u}_{2L}}}{100\gev} \right),
\label{eq:secbound}
\earr
where the first bound is correlated to the down-like squark mass and the second
bound to the up-like squark mass. The second bound on $\lam'_{123}$,
Eq.(\ref{eq:secbound}), is derived using $b$-quark asymmetry data and is
therefore more stringent.

The operator $L_1Q_2{\bar D}_1$ also contributes to atomic parity violation,
which leads to the slightly stricter bound \cite{vernon}
\beq
\lam'_{121} \lsim  0.26 \left( \frac{m_{{\tilde u}_{2L}}}{100\gev} \right) ,
\label{eq:bound4}
\eeq
depending on the $SU(2)$ doublet up-like squark mass. We see below, that
these bounds leave a significant discovery window at HERA.\footnote{There are
further relevant bounds on the couplings $\lam'_{111}$ from the non-observation
of neutrinoless double-beta decay \cite{doublebeta}, and on $\lam'_{122},\lam'_
{133}$, from bounds on neutrino masses \cite{rp1}, which are strongly model
dependent. They are the same magnitude as those given and we omit them in our
analysis.}

\medskip

There has also been an estimate of a collider bound on the operators $L_1Q_1
{\bar D}_1$ which could be obtained at CDF \cite{rp3}. This estimated bound is
a function of the slepton mass, not the squark mass and is therefore
complimentary to our results. Our bounds only very weakly depend on the
selectron mass, and we have chosen it to satisfy the estimated bound given in
Ref.\cite{rp3}.

\section{Single Squark Production at HERA}

Having identified the most promising process by which \rpv\ might be observed
at HERA, we now look more closely at this mechanism.

\subsection{ Squark Decays}
\label{sec:decrates1}
The squarks can decay to the 2-body final states shown in
Eqs.(\ref{eq:ddec},\ref{eq:udec}). For the \rpv\ decays the decay rate is given
by
\beq
\Gamma_1(m_q) = \frac{(\lam')^2}{16\pi} (1- \frac{m_q^2}{\msq^2})^2,
\eeq
where $\msq,m_q$ denote the mass of the initial squark and the final state
quark, respectively. We have neglected the mass of the electron. For the
cascade
decay to the photino, ${\tilde q}\ra q + \pho$, we obtain
\beq
\Gamma_2(m_q) = \frac{ e_q^2 e^2}{8\pi} \msq \, (1-\frac{m_q^2+M_\pho^2}
{\msq^2}) \, \mu^{1/2}(1,\frac{m_q^2}{\msq^2},\frac{M_\pho^2} {\msq^2})
\eeq
where $e_q$ is the charge of the squark in units of the electron charge,
$M_\pho$ is the photino mass  and the phase space function is given by
$\mu(x,y,z)=x^2+y^2+z^2-2xy-2xz-2yz$. In the limit of massless quarks we obtain
for the ratio of the cascade decay  to the \rpv\ decay
\barr
\frac{\Gamma_2}{\Gamma_1} &=&
\frac{2e^2e_q^2}{(\lam')^2}\left(1-\frac{M_\pho^2
}{ \msq^2}\right)^2\\ &\gsim & e_q^2 \left(1-\frac{M_\pho^2}{\msq^2}\right)^2
\left(\frac {100\gev} {\msq} \right)^2\left\{ \begin{array}{ll}  213, & L_1Q_1{
\bar D}_j, \\ 2.8, & L_1Q_2{\bar D}_{1,3}, \\ 0.95, & L_1Q_2{\bar D}_2,
\end{array} \right.
\earr
where we have used the Eqs.(\ref{eq:bound1}-\ref{eq:bound4}) to obtain the
inequalities. In the case of the operators $L_1Q_1{\bar D}_j$ we expect the
cascade decay to be dominant by more than an order of magnitude, whereas in the
other cases, due to the weakness of the existing bounds, the \rpv\ decay can be
comparable in size or even dominant.

The total decay widths of the squarks are given by
\barr
\Gamma_{tot}({\tilde d}_{jR}) &=& \Gamma_1(m_{ui}) + \Gamma_1(m_{dj}) +
\Gamma_2(m_{dj}),  \\
\Gamma_{tot} ({\tilde u}_{iL}) &=& \Gamma_1(m_{dj}) + \Gamma_2(m_{ui}).
\earr
The corresponding widths for the leptoquarks $S_1$ and ${\tilde R}_2$,
respectively, are obtained by setting $\Gamma_2\equiv0$.

\subsection{Photino Decay}
\label{sec:decrates}
The photino decays as shown in Figure~1. The differential decay rate for the
decay to the positron is given by
\barr
\frac{d\Gamma(\pho\ra e^+{\bar u}_{iL}d_{jR})}{dE_edE_d}&=& \frac{\alpha
\lambda'^2}{2\pi^2 M_\pho \sqrt{1 - m^2_d/E^2_d}} \left( e_e^2
\frac{{\bf(\pho.e)(d.u)}} {D^2_e} + e_u^2 \frac{{\bf (\pho.u)( e.d)}} {D^2_u}
\right. \nonumber
\\&&+e_d^2 \frac{{\bf (\pho.d)(e.u)}}{D^2_d} - e_ee_u
\frac{{\bf(\pho.e)(d.u)-(\pho.d)(e.u) +(\pho.u)(e.d)}}{D_eD_u} \nonumber\\
&&+e_de_e \frac{{\bf(\pho.d)(e.u)-( \pho.u)(e.d)+(\pho.e)(d.u)}}{D_eD_d}
\nonumber\\ &&\left. +e_ue_d\frac{{\bf(\pho.d)(e.u)
-(\pho.e)(d.u)+(\pho.u)(e.d)}}{D_uD_d}\right).
\earr
where $\alpha$ is the electromagnetic coupling, ${\bf\pho, e}$ and ${\bf d}$
are the 4-momenta of the photino, positron and the $d_j$-quark, respectively,
and ${\bf u = \pho-e-d}$ is the momentum of the final state up-like quark. $e_e
= -1$, $e_u=\frac{2}{3}$, $e_d=-\frac{1} {3}$ are the charges of the final
state particles in units of the absolute value of the electron charge. The
propagators squared are given by $D_e = m^2_\pho - m^2_{{\tilde e}} -
2{\bf\pho.e}$, $D_d = m^2_\pho - m^2_{{\tilde d}} - 2{\bf \pho.d}+ m^2_d$, and
$D_u= m^2_\pho - m^2_{{\tilde u}} - 2{\bf{\pho.u}} +m^2_u$. $E_e$ and $E_d$ are
the energy of the positron and the $d_j$ quark, respectively. The dot products
are easily expressed in terms of the masses and $E_e$ and $E_d$
\barr
{\bf\pho.e}&=& M_\pho E_e, \\
{\bf\pho.d}&=& M_\pho E_d,\\
{\bf\pho.u}&=& M_\pho^2-{\bf \pho.(e+d)},\\
{\bf e.d} &=& \half(M_\pho^2+m_u^2-m_d^2-2{\bf \pho.u}),
\earr
and we have neglected the mass of the electron.

For the other decay modes one obtains analogous formulae. Upon integrating
these decay rates  numerically we obtain the branching fraction for the decay
to the positron
\beq
BF(\pho\ra e^++2{\rm jets}) = \frac{\Gamma (\pho\ra e^++2{\rm jets}) }
{\Gamma_{tot}(\pho)} = 0.4378,
\label{eq:bfpho}
\eeq
where we have used $M_\pho=50\gev$, $\msq=150\gev$. The result only weakly
depends on these masses. For $M_\pho= 150\gev$ and $\msq=240\gev$ we obtain
0.4384.  This branching fraction was estimated to be $\quarter$ in Refs.
\cite{butter,rptop,dp}. If this result (\ref{eq:bfpho}) is included in Refs.
\cite{rptop,dp} the signals are enhanced by a factor of $\sim3.0$ leading to
stricter bounds for example in Eq.(\ref{eq:sqlle2},\ref{eq:sqlqd}) above.

\medskip

\subsection{Inclusive Reaction Cross Section}
At tree-level the electron-quark differential cross section for the process
(\ref{eq:one}) is given by
\barr
\frac{d{\hat\sigma}(e^-u_i\ra d_j\pho) }{d{\hat t}dx}&=&\frac{\lambda'^2e^2}
{32\pi\hat{s}^2}
\left(
e_d^2 \frac{{\hat s}({\hat s}-m^2_\pho-m^2_d)}{R_{\tilde d}^2}
+e_e^2\frac{({\hat t}-m^2_d)({\hat t}-m^2_{\pho})}{D_{\tilde e}^2}
+2e_de_e\frac{{\hat s}{\hat t}}{|R_{\tilde d}D_{\tilde e}|}
\right. \nonumber
\\
&&\!\!\!\!\!
\left.
+e_u^2\frac{({\hat u}-m^2_d)({\hat u}-m^2_{\pho})}{D_{\tilde
u}^2} + 2e_de_u\frac{{\hat s}{\hat u}} {|R_{\tilde d}D_{\tilde u}|}
+2e_ue_e\frac{{\hat t}({\hat u}-m^2_\pho m^2_d)}{|D_{\tilde u}D_{\tilde e}|}
\right),
\label{eq:xsec}
\earr
where $R_{\tilde d}^2=( {\hat s}-m_{\tilde d}^2)^2+ m_{\tilde d}^2
\,\Gamma_{tot
}^2({\tilde d})$, $D_{\tilde e}={\hat t}-m_{\tilde e}^2$, and $D_{\tilde u}={
\hat u}-m_{\tilde u}^2$. ${\hat s},{\hat t}$, and ${\hat u}$ are the parton
level Mandelstam variables. We have neglected the  masses of the incoming
particles but have included the masses in the final state. The parton level
cross section for the process given in Eq.(\ref{eq:two}) can be obtained from
the above formula through crossing. In order to obtain the electron-proton
cross
section from the parton level cross section we integrate over the product of
(\ref{eq:xsec}) and the appropriate $i-$th generation up-quark structure
function $u_i(x,T^2)$,
\beq
\frac{d{\sigma}(e^-P\ra d_j+ \pho+X)}{d{\hat t}}= \int
\frac{d{\hat\sigma}(e^-u_i\ra d_j\pho) }{d{\hat t}dx} \cdot u_i(x,T^2)\, dx,
\label{eq:totxsec}
\eeq
where $x$ is the usual Bjorken scaling variable and $T^2$ is the scale of the
transfered momentum squared. For $T^2$ we have chosen $\hat{s}$, by analogy
with photon-gluon fusion (PGF) processes. Using instead for instance Mandelstam
$t$ for the scale typically changes the cross section by less than $5\%$,
though for high squark masses the cross section can be enhanced by as much as
$15\%$. In Figure~3 we show the electron-proton cross section before
integration over $x$ and Mandelstam $t\,\,(= -Q^2_{DIS})$, so as a function of
$x$ at fixed $t(=49200~GeV^2)$.

\subsection{ Event Rates }

In the MSSM the renormalization group analysis usually leads to five nearly
mass degenerate squarks, with a possibly lighter sixth top squark. These
analyses should not be very much affected by a (small) \rpv\ Yukawa coupling.
Since we here do not consider top squark production we set $m_{{\tilde
d}_j}=m_{{\tilde  u}_i}=M_{SUSY}$. Furthermore, our results are insensitive to
to the selectron  mass and for simplicity and in order to satisfy the estimated
contraints of  Ref.\cite{rp3} we set $m_{\tilde e}=M_{SUSY}$ as well. The
dominant process is the s-channel squark production, so the squark and photino
masses will have a significant effect on event rates. Figure~4 shows the event
rates for a nominal year's running at HERA ($200~pb^{-1}$) as a function of
$M_{SUSY}$ for various photino masses. The kinematic region considered is;
\barr
4\, GeV^2 &<  Q^2_{DIS} &< 9.84\cdot 10^{4}\, GeV^2 ,\nonumber \\
10^{-4} &< x &< 0.9 .
\label{eq:kr}
\earr
As determined from Eq.(\ref{eq:xsec}) and seen in figure~3, the parton-level
cross section strongly peaks at the resonant squark masses and so this is where
most of the \rpv\ events are produced. The kinematic region (\ref{eq:kr})
includes the \rpv\ squark production peak for practically all accessible squark
masses and is chosen to minimise theoretical uncertainties from DIS physics.
The change in gradient in figure~4 at $M_{SUSY}$ of around $300~GeV$ is due to
the kinematic region used (\ref{eq:kr}) - at this point the squark resonant
peak falls outside the allowed $x$ range.

Figure~5 shows the cross section as a function of $\lambda'_{1ij}$ and for
fixed ${\tilde m}_{\tilde{q}}$, for three popular sets of structure function
parameterisations \cite{KMRS,MT,GRV}.

One of the main aims of the HERA experiments is to determine parton
distributions of the proton within the new kinematic regimes avaiable. The
present uncertainties in the valence quark distributions lie mainly in the low
$x$ $(< 10^{-2})$ and very high $x$ $(>0.4)$ regions. Using the resonant
approximation, $\msq^2\approx xs$, this corresponds to the wide range of squark
masses $30\ra 200\gev$. Therefore in the case of squark production from valence
quarks ($\lambda'_{11j} $) we find the uncertainties in the production rate due
to the uncertainties in the structure functions to be small, of the order of a
few percent. Due to greater uncertainties in the sea quark distributions,
uncertainties in the cross section for the $\lambda'_{12j}$ events can rise to
as much as fifty percent, though for most of the parameter space they remain at
around ten percent or less. These effects are illustrated in Figure~5. Plots
for varying squark mass were also generated and exhibit a similar range of
variation for the various parameterisations. In the case of $\lambda'_{11j}$,
the cross section is heavily dominated by interactions involing the $u$-valence
quarks in the proton, and so as can be seen, the differences between the cross
sections for $j=1,2,3$ are very small. The structure function parameterisations
were obtained from the PDFLIB package \cite{PDF}. For all subsequent studies in
this paper the KMRS B0 structure function set will be used.

\subsection{ Event Topology and Differential Cross Sections }

It is important to study \rpv\ event topologies in order to determine the best
cuts to separate signal from possible backgrounds at HERA. As discussed in
previous sections, for the process we are considering
\beq
e+q\ra (q')^*\ra q'+ {\tilde\gamma}, \quad {\tilde\gamma}\ra e^\pm+2\,jets
\eeq
the positron decay channel is seen as the most promising and thus positron
transverse momentum ($p_{Te}$) is an important quantity in our analysis.
Also as we expect both the photino and squarks to be relatively massive, thus
\rpv\ events generally have large values of total transverse energy ($E_T$).
Using a Monte-Carlo generator based upon the cross section and branching
fraction calculations described above, we now look in more detail at these
events.

Figure~6 shows plots of $d\sigma/dp_{Te}$ and $d\sigma/dE_T$ for three
different photino masses at $M_{SUSY} = 200~GeV$. Figure~7 shows the same
quantities for different values of $M_{SUSY}$ where the photino mass is fixed
at $50~GeV$. Note the log scale used here, made necessary by the strong
dependence of the cross section on $M_{SUSY}$. The Jacobian peak at around
$M_{SUSY}$ is evident in the transverse energy plots - particularly for the
lower photino masses. Also, at lower photino mass a second peak becomes visible
at around $M_{\pho}$; this is the ``Jacobian peak" from the photino decay for
those events in which the photino is emitted with low $p_T$.

It can be seen that $p_{Te}$ and $E_T$ are promising quantities to use in
selecting SUSY events, as was previously shown in Ref.\cite{butter}. We will
also make use of the total energy in the detector ($E_{TOTAL}$). The
corresponding cuts become potentially more powerful at higher SUSY masses,
which is also where the cross section falls off and thus efficiency is
especially important. In the next section we will study some of the possible
backgrounds to these processes, with particular reference to event
identification in the \zeus\ detector.

\section{ Detection of \rpv\ in \zeus\ }

The \zeus\ detector is positioned at one of the interaction points of HERA,
details of which can be found in Ref.\cite{zeus}. Here we restrict ourselves
to considering the Uranium Calorimeter (UCAL) and the Central Tracking Detector
(CTD). These are the two most important components when searching for high
$p_T$ positrons, the most promising signature for \rpv\ events.

\subsection{ Physics Background }

We have considered two types of physics background in detail. The first is
heavy quark production $via$ photon-gluon fusion (see, for example,
Ref.\cite{bbar}). These are generally low-$E_T$ events, but they have
relatively high cross sections at HERA and thus their high $E_T$ tail will
present a significant background. The most dangerous case can be expected to be
$b$-quark production. The $B^0$ thus produced can decay semi-leptonically to a
positron and a neutrino. The HERWIG generator \cite{HRW} was used to generate a
sample of these events. The cross section for $B^0$ production by this process
was estimated to be $2430~pb$ with the KMRS B0 set of structure functions. This
value can vary by as much as $30\%$ for other choices of structure function
parameterisations, due to the large uncertainties in the gluon distribution
within the proton.

The second background considered is $W^+$ production via the process
\beq
\gamma^* + q \ra W^+ + q',
\eeq
where the virtual photon is radiated from the electron and the $W^+$ can decay
to give a positron. Detailed studies of this process at HERA were presented in
\cite{zep1,zep2}. A sample of these event was generated with the generator used
in the first of these references. The cross section for this process multiplied
by the branching fraction for the ($W^+ \ra e^+  \nu_e$) was estimated
to be $78~fb$, varying by as much as $40\%$ for different choices of the
structure function.

\subsection{ \lq Fake\rq\ Background }
\lq Fake\rq\ backgrounds are those caused not by genuine positron production,
as in the processes above, but by effects inside the \zeus\ detector faking a
high $p_T$ positron signal. These include
\begin{enumerate}
\item Misidentification of the sign of the curvature (and thus charge) of high
$p_T$ electrons in the tracking detector;

\item Overlap of hadrons (particularly $\pi^+$) with photons, providing a
positive track in the CTD pointing to an electromagnetic cluster in the UCAL.
\end{enumerate}

In order to investigate the first effect, a sample of high $Q^2$ deep inelastic
scattering neutral current events were generated, using the program LEPTO
\cite{LEPTO}. These events provide a source of high $p_T$ electrons.  The
kinematic region over which these events were generated was
\barr
300\,GeV^2 &< Q^2 &< 9.84\cdot 10^{4}\, GeV^2 ,\nonumber \\
10^{-4} &< x &< 0.9.
\earr
The cross section within this region is $1350~pb$ for the KMRS B0 set of
structure functions. This value changes by around $10\%$ for the various
choices
of structure function.

Studying the second of these effects requires a detailed understanding of the
\zeus\ detector and its environment. Previous studies \cite{butter} indicate
that these effects can be reduced to a level below that of the other
backgrounds mentioned.

\subsection{ Efficiencies and Resolution }

In order to gain some knowledge of the effectiveness of experimental cuts and
thus the possible discovery limits for \rpv\ in \zeus\, samples of \rpv\ and
background were generated corresponding to at least $20~pb^{-1}$ of HERA
luminosity. These events were then passed through a simple simulation of the
\zeus\ detector, consisting of smearing of particle momenta and energies by
nominal detector resolutions, followed by geometrical cuts for detector
acceptance. The calorimeter resolutions, taken from Ref.\cite{zeus}, were
$\frac{17\%}{\sqrt{E}}$ for the electromagnetic calorimeter, and $\frac{35\%}
{\sqrt{E}}$ for hadrons. An angular resolution of $10~mrad$, which is quoted in
\cite{zeus} as the worst resolution anywhere within the angular acceptance of
the \zeus\ calorimeter, was also applied. The CTD resolutions used are
summarised in Table~2. Where a charged particle entered the CTD, information
about its angle was taken from the CTD. All energies were taken from the
calorimeter. CTD momentum information was only used to determine the sign of
the charge for positron candidates.

The events were generated at vertices $x = y = 0, z = -20cm \ra +20cm$ assuming
a uniform distribution, where the origin is defined to be the centre of the
\zeus\ detector.

The \rpv\ events were generated for various squark masses and couplings at a
photino mass of $M_{\pho} = 50\,GeV$, for the photino decays to $e^-$ or $e^+$
and two jets. Although we attempt to identify only events with positrons in the
final state, detector resolution will mean that there will be some
misidentification of positrons as electrons, and thus some mixing between
events exhibiting these two types of decay.

Since the event characteristics change significantly with squark mass, we have
developed two sets of cuts, one optimised for the lower range of squark masses
(we have considered squark masses down to $100\,GeV$) and the other for squark
masses at the higher limit of observation at HERA.

For the limit of small $\lambda'_{1ij}$, which is only accessible at low squark
mass, the cuts rely heavily on the details of the positron signal and must be
fairly elaborate. The series of cuts found to be optimum in this case, and the
corresponding efficiencies for $M_{SUSY} = 100~GeV, M_{\pho} = 50~GeV$ and
$\lambda'_{111} = 0.02$  are shown in Table~3. Figure~8 shows the total energy
distribution for background and signal before and after these cuts, for the
same parameters. The cuts which were found to be optimum for the limit of high
squark mass are summarised in the Table~4. They can be less complex than the
cuts for lower squark masses, due to the large amounts of transverse energy
present in \rpv\ events at high squark masses. Because the $E_T$ cut is very
powerful, the cut on the transverse momentum of the positron can be relaxed
somewhat. Also shown are the number of events passing each of these cuts, for
$M_{SUSY} = 240~GeV, M_{\pho} = 50~GeV$ and $\lambda'_{111} = 0.07$. Figure~9
shows total energy distributions for background and signal events before and
after these cuts, for $M_{SUSY} = 280~GeV, M_{\pho} = 50~GeV$ and
$\lambda'_{111} = 0.084$

\section{Results}

Figure~10 shows the contour in $\lam'_{1ij}$ space corresponding to five events
after $200~pb^{-1}$ luminosity and before any detector acceptance or cuts are
applied to the data. Also shown are the existing limits applying for any or all
of these couplings, as detailed in sections 4.2 and 5.1. Values to the right of
these lines are already ruled out for the couplings $\lam'_{1ij}$ as indicated
on the plot.

In order to explain our procedure for obtaining the possible discovery reach at
\zeus\, take as an example the most promising case of $\lam'_{11j}$. The
existing limit given in Eq.(\ref{eq:bound1}) defines a curve in mass-coupling
space. We generated \rpv\ events at various masses and couplings spaced along
this curve. Applying detector smearing, followed by the two sets of cuts from
Tables~3 and~4 to these events, we obtain the event numbers and
signal/background ratios shown in Figure~11. Where the background is so low
that none of the Monte-Carlo background events passed the cuts, the background
value corresponding to a one sigma upward fluctuation was used in calculating
the ratio. It can be clearly seen that the first set of cuts is most effective
at the lower squark masses, whereas the second set is optimised for masses
around $240\gev$. The error bars indicate the statistical errors due to the
size of the Monte-Carlo samples generated.

Applying the discovery limit criteria:

\begin{itemize}
\item Five or more \rpv\ events surviving our cuts per nominal HERA year ($
200pb^{-1}$).
\item A signal/$\sqrt{\rm background}$ ratio of at least three.
\end{itemize}
to these plots we calculate that \rpv\ should be observable in the \zeus\
detector
up to a squark mass of
\beq
M_{\tilde{q}} \lsim 270 \,GeV,
\eeq
for $\lam'_{11j} = 0.081$. For $M_{\tilde{q}} = 100\,GeV$, \rpv\ should be
observable for
\beq
\lam'_{11j} \gsim 0.0053
\eeq
Carrying out a similar procedure for the $\lam'_{12i}$ cases we calculate that
\rpv\ should be observable for
\barr
M_{\tilde{q}} \lsim 205~GeV, \quad \lam'_{121} = 0.53,  \\
M_{\tilde{q}} = 100 GeV, \quad \lam'_{121} \gsim 0.013, \\
M_{\tilde{q}} \lsim 205~GeV, \quad \lam'_{122} = 0.92,  \\
M_{\tilde{q}} = 100 GeV, \quad \lam'_{122} \gsim 0.015, \\
M_{\tilde{q}} \lsim 175~GeV, \quad \lam'_{123} = 0.46,  \\
M_{\tilde{q}} = 100 GeV, \quad \lam'_{123} \gsim 0.023 .
\earr

It is worth pointing out that it should be possible to extend an analysis by
Doncheski and Hewett \cite{doncheski} on the effects of virtual leptoquarks
at HERA to \rpv\ and squarks and thus obtain bounds on much higher mass
squarks.

\section{Conclusions}

We have presented a specific calculation showing a large expected reach at HERA
into kinematic regions where R-Parity (\rpv) could be discovered.

\rpv\ could be discovered for $\lambda'_{11j}$ as low as $5.8 \times 10^{-3}$
or for squark masses as high as $270\,GeV$ at $\lambda'_{11j} = 0.081$ after
the first $200~pb^{-1}$ of HERA luminosity. For the other couplings, {\it i.e.}
$\lambda'_{12i}$, the reach in squark mass is somewhat lower, but the weaker
existing limits on these couplings mean that there is still a substantial
window for discovery at HERA. These results will hold for photino mass less
than squark mass by a few tens of $\gev$ or more.

These limits compare with a possible exclusion limit for $S_0$ leptoquarks
after $100~pb^{-1}$ of $275\,GeV$ at a similar value of coupling $\lam_L$,
obtained in Ref.\cite{sirois}. We conclude that HERA is a very promising
experiment for the discovery of  supersymmetry with broken $R_p$ and we eagerly
await the results of the present run.

\medskip

\noindent {\bf {\large Acknowledgements}}

Thanks are due to Dieter Zeppenfeld, for helpful discussions and use of his
computer program for $W^+$ production at HERA. We thank Neville Harnew for
making many valuable suggestions. We also thank Manuel Drees for discussions of
the MSSM associated pair production at HERA.

\appendix

\section{Table of Dominant \rpv-Signals}

Here we explain in detail Table 1 which summarizes the \rpv\ signals at HERA.
The analogous table for hadron coliders has been given in Ref.\cite{rphadron}
and we employ the same notation here. We first consider the MSSM pair
production
mechanisms in the first three horizontal boxes, before we come to the single
SUSY production via \rpv\  in the bottom horizontal box. The signals are
determined out of the combination of the SUSY production mechansim and the
\rpv\ decays as indicated in Eqs(\ref{eq:signal1},\ref{eq:signal2}). So we in
turn discuss the production and decay mechanisms.

At HERA the dominant pair production mechanisms in the MSSM are
\barr
e^-+q& \ra& {\tilde e}^- + {\tilde q} \\
e^-+q &\ra& {\tilde \nu}_e + {\tilde q}'
\earr
as well as
\beq
e^- + (\gamma^*,Z^{0*}) \ra {\tilde e}^- + \lsp,
\eeq
where $\gamma^*,Z^{0*}$ denote the virtual neutral gauge bosons.
This latter process has been discussed in \cite{manueldieter} for the case of
conserved \rp. The cross section is relatively low and we have not
 made a numerical estimate for the case of \rpv\ decays. However, for
completeness we include the dominant signals.

The respective
pair produced SUSY particles ${\tilde e}^- {\tilde q},\,{\tilde \nu}_e {\tilde
q}',$ and $ {\tilde e}^- \lsp,$ are listed in the first three boxes on the left
of the first column. We  distinguish the signals according to which
class the single dominant operator belongs ({\it cf.} the assumptions in
Section
3).

The dominant decays of the scalar fermions were determined in
Ref.\cite{rphadron} and include two different modes. They either decay via a
conventional SUSY coupling
\barr
{\tilde e}& \ra & e +\lsp \\
{\tilde q} & \ra & q + \lsp,
\earr
or via an \rpv-operator
\barr
{\tilde e} &\ra & {\cal O}({\tilde e})\equiv \left\{ \begin{array}{ll}
(e,\mu,\tau)+\nu, & LL{\bar E} \\
2\,\, jets, & LQ{\bar D} \\
-, & {\bar U}{\bar D}{\bar D} \end{array} \right. \\
{\tilde q} &\ra & {\cal O}({\tilde q})\equiv \left\{ \begin{array}{ll}
-, & LL{\bar E} \\
(e,\nu)+1\,\, jet, & LQ{\bar D} \\
2\,\, jets & {\bar U}{\bar D}{\bar D} \end{array} \right.
\earr
We have introduced the notation ${\cal O}({\tilde f})$ to collectively
denote the decay
products of a scalar fermion via an \rpv\ operator. The blanks indicate cases,
where the sfermion does not couple to the operator at tree level. In
Ref\cite{rphadron} it was shown that these \rpv\ decays are never dominant
for operators $LL{\bar E}$ due to the existing restrictve laboratory bounds
\cite{vernon}. For the operators $L_iQ_j{\bar D}_k$ they can only dominate
for higher generational indices. The conditions determining which decays are
dominant are discussed in detail in \cite{rphadron}. Clearly the decay products
and therefore ultimately the signals depend on the single dominant operator
under consideration.

In \rpv\ the LSP also decays within the detector via the diagram shown in
Figure
1, provided the constraint \ref{eq:lam} is satisfied. The resulting decay
products are
\beq
\lsp\ra \left\{ \begin{array}{ll}
e_i^\pm e_k^\mp\nu_j, & L_iL_j {\bar E}_k, \\
(e_i^\pm,\nu) + 2\,\,jets, & L_iQ_j{\bar D}_k, \\
3\,\,jets, & {\bar U}_i {\bar D}_j {\bar D}_k,
\end{array}
\right.
\label{eq:lspdecays}
\eeq
where again the decay products depend on the single dominant operator. As
discussed in Ref.\cite{dawson,rphadron}, in special cases involving only the
operators $L_iQ_j{\bar D}_j$ (note the last two indices are identical) the
LSP can also dominantly decay via the radiative process $\lsp\ra\gamma+\nu$.
This is discussed in more detail in \cite{rphadron}.

The dominant and most promising final state signals for a given dominant
operator are summarized in the last three columns. In order to determine the
complete final state we must collect the decay products of all produced SUSY
particles. As an example consider the first line in the first horizontal box.
The production mechanism is $e^-q\ra{\tilde e}^-{\tilde q}$. This is followed
by
the in most cases dominant cascade decay of the scalar fermions:
\beq
{\tilde e}^-{\tilde q}\ra e^-q+2\lsp
\eeq
which is the entry in the first column and first row of the first horizontal
box.

For a dominant operator $L_iL_j{\bar E}_k$ the LSP decay is given in
Eq.(\ref{eq:lspdecays}). The
index $k$ fixes one of the final state {\it charged} leptons to be either
$e,\mu,$ or $\tau$. Since the first two indices are anti-symmetric, at
least one of them is unequal to 3. Thus of the possible decay products
relating to the first two indices: $(e,\mu,\tau)+\nu$, we can choose as
the promising case: $(e,\mu)+\nu$. The combined possibilities we write as
\beq
(ee,e\mu,\mu e,\mu\mu,\tau e,\tau\mu)+\nu\equiv(e,\mu,\tau)^\pm\otimes(e,\mu)
+\nu
\eeq
The final state neutrino results in $\ptmis$ and since we have two LSPs the
total signal is
\beq
e^-q+\ptmis+ 2 (e,\mu,\tau)^\pm \otimes (e,\mu)^\mp
\eeq
which is the entry given in the second column and first row of the first
horizontal box in the table.

The other entries in the table are determined by considering the other decay
modes of the sfermions and the LSP.
The second row in the first horizontal box gives the signal for the special
radiative decay of the LSP: $\lsp\ra\gamma+\nu$. The third row is for the case
where only the squark decays via a \rpv\ operator and the fourth row of the
first horizontal box summarizes the signal where both sfermions decay through
an \rpv\ operator.

The signals for the charged current process $e+q\ra{\tilde\nu}+q'$ as well as
the electron-photon scattering $e+(\gamma^*,Z^{0*})\ra{\tilde e}+\lsp$ are
summarised in the second and third horizontal boxes.

One immediate conclusion is that the operators $UDD$ do not lead to any
promising signals at HERA. Whereas the operators LLE lead to many charged
leptons which should make them easy signals to distinguish.

In the last horizontal box we have summarized the signals for the case of
single SUSY production at HERA, which is discussed in detail in this paper.
 The production mechanism is the resonant
s-channel diagram given in Figure 2a. The squark decays are discussed in detail
in Section 4.1. The production mechanism is restricted to the operators
$LQ{\bar D}$ and the final state signals are summarised in column three at the
bottom of the Table.

\medskip

\noindent {\bf Note Added:} Just before completion we obtained a preprint by
J. Hewett \cite{joanne2} which contains related work.

\pagebreak

\begin{table}
\begin{center}
\begin{tabular}{||c||c|c|c||} \hline
{\em Quantity}           &  \multicolumn{3}{||c||}{\em Polar Angle} \\
\cline{2-4}
{\em  }            & $90^o\ra 45^o$ & $45^o\ra 22^o$ & $22^o\ra 15^o$ \\ \hline
$1/Momentum$            & $0.0018$       & $0.0012$       & $0.012$ \\
$(GeV/c)^{-1}$          & $+0.0029/p$    & $+0.0032/p$    & $+0.011/p$\\ \hline
$\theta$                & $0.0012$       & $0.00061$       & $0.0014$ \\
$(radians)$             & $+0.0015/p$    & $+0.0015/p$    & $+0.0026/p$\\
\hline
$\phi$                  & $0.00059$      & $0.00053$       & $0.0067$ \\
$(radians)$             & $+0.0015/p$    & $+0.0023/p$    & $+0.0050/p$\\
\hline
\end{tabular}
\caption{Angle and momentum dependence of CTD resolutions used.}
\end{center}
\end{table}

\begin{table}
\begin{center}
\begin{tabular}{||c|c||c|c|c|c||} \hline
{ Variable}         & { Cut }   & \multicolumn{4}{||c||}{Events Remaining} \\
\cline{3-6}
{ used }    & {     }   & {\rpv}  & {$b\bar{b}$} & {NC DIS} & {$W^+$} \\ \hline
{None}            & $-$ & $1440$ & $5.1\times 10^5$ & $2.6\times 10^5$ & $16$\\
$p_{Te} (GeV/c)$        & $>7$  & $537$  & $185$      & $123$   & $12$\\
$E_T (GeV/c^2)$         & $>40$    & $527$      & $139$      & $123$   & $10$\\
$E_{TOTAL} (GeV/c^2)$   & $>40$    & $527$      & $139$      & $113$   & $8$\\
$Missing E_T (GeV/c^2)$ & $<30$   & $508$     & $124$    & $96$   & $2$\\
\hline
\end{tabular}
\caption{Cuts (1) and Efficiencies, $M_{SUSY}=100\,GeV$ }
\end{center}
\end{table}

\begin{table}
\begin{center}
\begin{tabular}{||c|c||c|c|c|c||} \hline
{ Variable}         & { Cut }   & \multicolumn{4}{||c||}{Events Remaining} \\
\cline{3-6}
{ used }            & {     }   & {\rpv}  & {$b\bar{b}$} & {NC DIS} & {$W^+$}
\\ \hline
{None}              & $-$  & $152$ & $5.1\times 10^5$ & $2.6\times 10^5$ &
$16$\\
$p_{Te} (GeV/c)$     & $>2$   & $63$  & $12700$      & $1550$   & $13$\\
$E_T (GeV/c^2)$      & $>120$ & $56$  & $0$          & $26$     & $0.7$\\
$E_{TOTAL} (GeV/c^2)$ & $>450$ & $56$  & $0$          & $0$     & $0.2$\\
\hline
\end{tabular}
\caption{Cuts (2) and Efficiencies, $M_{SUSY}=240\gev$ }
\end{center}
\end{table}

\pagebreak

\begin{figure}
\caption{\label{f:1} LSP decay }
\end{figure}

\begin{figure}
\caption{\label{f:2} a) Single Squark Production  b) Leptoquark-like Decay}
c) Cascade Decay
\end{figure}

\begin{figure}
\caption{\label{f:3} \rpv\ ep Differential Cross Section  }
\end{figure}

\begin{figure}
\caption{\label{f:4} \rpv\ events rates at HERA }
\end{figure}

\begin{figure}
\caption{\label{f:5} \rpv\ ep Cross Section for different structure function
parameterisations }
\end{figure}

\begin{figure}
\caption{\label{f:6} \rpv\ Differential Cross Sections for various photino
masses.}
\end{figure}

\begin{figure}
\caption{\label{f:7} \rpv\ Differential Cross Sections for various squark
masses.}
\end{figure}

\begin{figure}
\caption{\label{f:8} Total Energy distributions for events containing
positrons}
a) Before cuts, and b) after application of cuts in table~3.
$M_{SUSY} = 100~GeV, M_{\pho} = 50~GeV$ and $\lambda'_{111} = 0.02$
\end{figure}

\begin{figure}
\caption{\label{f:9} Total Energy distributions for events containing
positrons}
a) Before cuts, and b) after application of cuts in table~4.
$M_{SUSY} = 280~GeV, M_{\pho} = 50~GeV$ and $\lambda'_{111} = 0.084$
\end{figure}

\begin{figure}
\caption{\label{f:10} \rpv Cross Section contours and existing bounds.}
\end{figure}

\begin{figure}
\caption{\label{f:11} Signal/Background at discovery limit}
\end{figure}

\end{document}